\documentclass[twocolumn,prb,amsmath,amssymb,amsfonts,superscriptaddress,floatfix,aps,showpacs,notitlepage]{revtex4-1}

\usepackage{graphicx}
\usepackage{epstopdf}
\usepackage{dcolumn}
\usepackage{bm}
\usepackage{rotating} 
\usepackage{color}
\usepackage{subfigure}
\usepackage{natbib}
\usepackage{enumitem}
\usepackage[Symbol]{upgreek}
\usepackage[symbol*]{footmisc}
\usepackage{lipsum}
\usepackage{amsmath}                    
\usepackage{amssymb}                    

\usepackage{textcomp}		
\usepackage{tabularx}
\usepackage{float}		

\begin{document}
\newcommand{\oscar}[1]{\textcolor{red} {#1}}
\renewcommand{\thetable}{\arabic{table}}

\title{\emph{Ab-initio} study of the Coulomb interaction in Nb$_{x}$Co clusters: \\ Strong on-site versus weak non-local screening}

\date{\today}

\author{L. Peters}
\email{L.Peters@science.ru.nl}
\affiliation{Institute for Molecules and Materials, Radboud University 
Nijmegen, NL-6525 AJ Nijmegen, The Netherlands}

\author{E. \c{S}a\c{s}{\i}o\u{g}lu}
\email{ersoy.sasioglu@physik.uni-halle.de}
\affiliation{Institut f\"ur Physik, Martin-Luther-Universit\"at Halle-Wittenberg, 06120 Halle (Saale) Germany}

\author{I. Mertig}
\affiliation{Institut f\"ur Physik, Martin-Luther-Universit\"at Halle-Wittenberg, 06120 Halle (Saale) Germany}

\author{M. I. Katsnelson}
\affiliation{Institute for Molecules and Materials, Radboud University Nijmegen, NL-6525 AJ Nijmegen, 
The Netherlands}

\begin{abstract}
By means of \textit{ab-initio} calculations in conjunction with the random-phase approximation 
(RPA) within the full-potential linearized augmented plane wave method we study the screening 
of the Coulomb interaction in Nb$_{x}$Co ($1 \leq x \leq 9$) clusters. In addition, these results 
are compared with pure bcc Nb bulk. We find that for all clusters the onsite Coulomb interaction in RPA
is strongly screened whereas the inter-site non-local Coulomb interaction is weakly screened and for 
some clusters it is unscreened or even anti-screened. This is in strong contrast with pure Nb bulk, 
where the inter-site Coulomb interaction is almost completely screened. Further, constrained RPA
calculations reveal that the contribution of the  Co 3\emph{d} $\rightarrow$ 3\emph{d} channel to 
the total screening of the Co 3\emph{d} electrons is small. Moreover, we find that both the onsite and 
inter-site Coulomb interaction parameters decrease in a reasonable approximation linearly with the cluster 
size and for clusters having more than 20 Nb atoms a transition from 0D to 3D screening is expected to 
take place.

\end{abstract}

\maketitle


\section{Introduction}

The interest in the field of clusters is growing due to the increasing demand for nano-technology. 
Besides the relevance for technology, clusters are also fundamentally very interesting. They behave 
in general very different from their bulk counterpart. Also their electronic and magnetic properties 
can drastically change by just adding or removing one atom.\cite{jena,chris,lars1,lars2} For example, 
in a recent work on Nb$_{x}$Co clusters it is demonstrated that Nb$_{5}$Co and Nb$_{7}$Co are non-magnetic, 
while  Nb$_{4}$Co and Nb$_{6}$Co are strongly magnetic.\cite{larsrem} The physical origin of this behavior 
can be traced back to the drastic change of the electronic structure as a function of cluster size.

The study of bimetallic clusters offers a broader playground than for pure clusters. This has resulted 
in a number of intriguing observations.\cite{larsrem,lars3,lars4,pram,pram2,tay} For example, from an 
anion photo-electron spectroscopy study on bimetallic Nb$_{x}$Co clusters, it was observed that for $x=6$ 
due to the addition of one Co atom to the Nb$_{x}$ host, the electronic structure resembles that of a typical bulk semi-conductor.\cite{pram} Therefore, this cluster was then proposed as a candidate for semiconductor materials.

From the theoretical side the bond properties and electronic structure of (NbCo)$_{x}$ clusters has 
been investigated by means of relativistic density functional theory (DFT).\cite{wang} Further, the 
geometry, stability and electronic properties of neutral and anionic Nb$_{x}$Co clusters is compared 
with pure Nb$_{x}$ clusters within a DFT study.\cite{hong} Recently a combined theoretical 
and experimental investigation has been performed on Nb$_{x}$Co clusters.\cite{larsrem} In this work 
the geometry is obtained from a comparison of experimentally and theoretically obtained vibrational 
spectra. With the geometry established the electronic structure is investigated in order to explain 
the magnetic properties obtained from magnetic deflection experiments.

To our knowledge a systematic assessment of screening and correlation effects in Nb$_{x}$Co clusters does 
not exist. This information is crucial in order to obtain a proper fundamental understanding of the system. 
Namely correlation effects among the electrons inhibit in general an exact solution. Therefore, approximate 
methods are required in practice. The choice of a suitable approximate method requires knowledge of the 
effective Coulomb interaction in the system. More precisely, the gradient of the effective Coulomb 
interaction is of importance.\cite{mot1,mot2} A very small gradient means that the effective Coulomb 
interaction is merely constant, while a very large gradient indicates a purely local effective Coulomb 
interaction. In the former case a mean-field treatment, i.e. single-particle approach, is probably a good 
choice, while for the latter it might be the (generalized) Hubbard model.

The aim of the present work is the \textit{ab-initio} determination of the Coulomb interaction for NbCo to Nb$_{7}$Co 
and Nb$_{9}$Co clusters. Besides being fundamentally interesting, such information is crucial to select an 
adequate theoretical method for a further investigation of the system. The geometries of Nb$_{3}$Co to 
Nb$_{7}$Co and Nb$_{9}$Co are well established from a comparison of theoretically and experimentally 
obtained vibrational spectra.\cite{larsrem} In addition NbCo and Nb$_{2}$Co are considered, since the number 
of isomers is very small. All these  clusters are known to be magnetic except Nb$_{5}$Co and Nb$_{7}$Co, 
which are non-magnetic. By employing the full-potential linearized augmented plane wave (FLAPW) method 
using Wannier functions in conjunction with the random-phase approximation (RPA)\cite{ferd1,miy1,ers1}, 
it is found that in these clusters the onsite Coulomb interaction in RPA is well screened, while the 
inter-site Coulomb interactions are barely screened. Interestingly for NbCo the inter-site interaction is 
unscreened, while for Nb$_{4}$Co even anti-screening occurs. The important consequence being that the screened 
Coulomb interaction is almost constant throughout the clusters. For completeness these results are compared 
with pure Nb bulk for which only the onsite Coulomb interaction is appreciable, while the inter-site Coulomb 
interactions are almost completely screened. Moreover, our constrained RPA calculations reveal that the 
Co 3\emph{d} $\rightarrow$ 3\emph{d} channel only plays a minor role in the screening of the onsite 
Coulomb interaction of the Co 3\emph{d} electrons. Finally, we find that both the onsite and inter-site 
Coulomb interaction parameters decrease in a reasonable approximation linearly with the cluster size and 
for clusters having more than 20  Nb atoms a transition from 0D to 3D screening is expected to take place. The 
rest of the paper is organized as follows. The method and computational details are presented in 
Section\,\ref{section-2}. Section\,\ref{section-4} deals with the results and discussion and finally in 
Section\,\ref{section-5} we give the conclusions.

\section{Method and computational details} 
\label{section-2}

In this work the screening of the Coulomb interaction in Nb$_x$Co clusters is calculated by means of the 
ab-initio random phase approximation (RPA) method. The non-interacting reference system required for this method 
comes from a preceding DFT calculation. In the following we shortly explain the RPA method and for details we 
refer to Ref.~\onlinecite{lars3}. The screened Coulomb interaction is defined as 
\begin{equation}
W(\boldsymbol{r},\boldsymbol{r}',\omega)=\int d\boldsymbol{r}''  \epsilon^{-1}(\boldsymbol{r},\boldsymbol{r}'',\omega) v(\boldsymbol{r}'',\boldsymbol{r}'),
\label{fullysw}
\end{equation}
where $\epsilon(\boldsymbol{r},\boldsymbol{r}'',\omega)$ is the dielectric function and $v(\boldsymbol{r}'',\boldsymbol{r}')$ 
is the bare Coulomb interaction potential. Since an exact expression for the dielectric function is not accessible, 
an approximation is required. In the RPA the dielectric function is approximated by
\begin{equation}
\epsilon(\boldsymbol{r},\boldsymbol{r}',\omega)=\delta(\boldsymbol{r}-\boldsymbol{r}')-\int d\boldsymbol{r}'' v(\boldsymbol{r},\boldsymbol{r}'')P(\boldsymbol{r}'',\boldsymbol{r}',\omega),
\label{rpadiel1}
\end{equation}
where the polarization function $P(\boldsymbol{r}'',\boldsymbol{r}',\omega)$ is given by
\begin{equation}
\begin{gathered}
P(\boldsymbol{r},\boldsymbol{r}',\omega)=\\
\sum_{\sigma} \sum_{\boldsymbol{k},m}^{occ} \sum_{\boldsymbol{k}',m'}^{unocc} \varphi_{\boldsymbol{k}m}^{\sigma}(\boldsymbol{r}) \varphi_{\boldsymbol{k}'m'}^{\sigma*}(\boldsymbol{r}) \varphi_{\boldsymbol{k}m}^{\sigma*}(\boldsymbol{r}') \varphi_{\boldsymbol{k}'m'}^{\sigma}(\boldsymbol{r}')  \\
\times\Bigg[ \frac{1}{\omega-\Delta_{\boldsymbol{k}m,\boldsymbol{k}'m'}^{\sigma}} - \frac{1}{\omega+\Delta_{\boldsymbol{k}m,\boldsymbol{k}'m'}^{\sigma}} \Bigg].
\end{gathered}
\label{rpapol1}
\end{equation}
\newline
Here $\Delta_{\boldsymbol{k}m,\boldsymbol{k}'m'}^{\sigma}=\epsilon_{\boldsymbol{k}'m'}^{\sigma}-\epsilon_{\boldsymbol{k}m}^{\sigma}-i\eta$ 
with $\epsilon_{\boldsymbol{k}m}^{\sigma}$ the single particle Kohn-Sham eigenvalues obtained from DFT and $\eta$ a positive infinitesimal. 
Further, the $\varphi_{\boldsymbol{k}m}^{\sigma}(\boldsymbol{r})$ are the single particle Kohn-Sham eigenstates with 
spin $\sigma$, wavenumber $\boldsymbol{k}$ and band index $m$. The tags $occ$ and $unocc$ above the summation symbol indicate that the 
summation is respectively over occupied and unoccupied states only.

The Eqs.~(\ref{fullysw}), (\ref{rpadiel1}), and (\ref{rpapol1}) constitute what is called the RPA of the dynamically 
screened Coulomb interaction. It is also possible to exclude certain screening contributions from Eq.~(\ref{rpapol1}), 
which is referred to as \textit{constrained} RPA (cRPA). In this work the screening of the Coulomb interaction for the Co 
3\textit{d} electrons and Nb 4\textit{d} electrons are investigated. One could for example exclude the screening contribution 
coming from the Co 3\textit{d} states to obtain insight in their contribution to the total screening. More details on 
the method used in this work to exclude certain screening contributions can be found in Ref.~\onlinecite{ers1}. Note that recently cRPA has become a very popular method to calculate Coulomb interaction parameters for different classes of materials.~\cite{Kotani,Schnell,Solovyev,Cococcioni,cRPA1,Zhang,Hunter}

The DFT calculations, providing the input of Eq.~\ref{rpapol1}, are performed with the FLEUR code. This code is 
based on a FLAPW implementation.\cite{fleur} All calculations are performed with an exchange-correlation functional 
in the generalized gradient approximation (GGA) as formulated by Perdew, Burke and Ernzerhof (PBE).\cite{gga} Further, all calculations are without spin orbit coupling. As will be demonstrated the effect of screening is on the $eV$ energy scale, while for the Co 3\textit{d} and Nb 4\textit{d} electrons the spin orbit coupling strength is at least an order of magnitude smaller. In addition it will be shown (Table~\ref{tabuco} and Fig.~\ref{fig3}) that the contribution of these electrons to the screening is small with respect to the other electrons, the Co 4\textit{sp} and Nb 5\textit{sp} electrons. For such extended states the spin orbit coupling strength is even smaller than for the Co 3\textit{d} and Nb 4\textit{d} electrons. Therefore, spin orbit coupling effects are expected to be small for the consideration of effective interactions. Since FLEUR is a {\bf k}-space code, a supercell approach was employed for the cluster calculations, with a large vacuum between clusters that were repeated in a periodic lattice. In order to prevent the interaction between clusters of different unit cells we performed tests for different unit cell sizes. We found that for a large unit cell of 12~\AA ~dimensions the results are converged to within a few percent. Therefore, this unit cell size is used for our calculations. Further, for the cluster calculations 
the cutoff for the plane waves is 4.0~Bohr$^{-1}$, $l_{cut}=8$ and the $\Gamma$ point is the only {\bf k}-point 
considered. For the calculations of bulk bcc Nb  we use the same parameters with a $20 \times  20 \times 20$ 
\textbf{k}-point mesh and experimental lattice parameter of 3.3~\AA~of the bcc lattice.  The ground state 
geometric and magnetic structure of the Nb$_{3}$Co to Nb$_{7}$Co and Nb$_{9}$Co clusters is obtained from 
Ref.~\onlinecite{larsrem} (see also Fig.~\ref{figclus}). More precisely, the geometries and magnetic structure 
are obtained from a comparison of calculated and measured vibrational spectra. Structures of NbCo, Nb$_{2}$Co 
and Nb$_{8}$Co were not addressed in Ref.~\onlinecite{larsrem}. Since the structure for Nb$_{8}$Co is unclear 
due to the many possible isomers, we will only address NbCo and Nb$_{2}$Co in addition. In order to obtain the 
ground state geometry of NbCo and Nb$_{2}$Co we performed the ATK-DFT calculations \cite{ATK-DFT} using the 
GGA-PBE exchange-correlation functional \cite{ATK-GGA} and the SG15-Medium combination of norm-conserving 
pseudopotentials and LCAO basis sets.\cite{Stradi,SG15} The total energy and forces have been converged at least 
to ∼ $10^{-4}$ eV and 0.01 eV/\AA, respectively.

The DFT calculations are used as an input for the SPEX code to perform RPA and cRPA calculations for the screened 
and partially screened (Hubbard $U$) Coulomb interaction.\cite{spex} The SPEX code uses the Wannier90 library to 
construct the maximally localized Wannier functions.\cite{wan90,wan902} For this construction we used per spin channel six 
states per Co atom, i.e. five 3$d$ states and one 4$s$ state, and nine states per Nb atom, five 4$d$ states, 
one 5$s$ state and three 5$p$ states. More precisely, the maximally localized Wannier functions are used to 
project the screened (bare) Coulomb interaction of Eq.~(\ref{fullysw}) on,
\begin{equation}
\begin{gathered}
U_{in_{1},jn_{3},in_{2},jn_{4}}^{\sigma_{1},\sigma_{2}}(\omega)= \\
\int \int d\boldsymbol{r}d\boldsymbol{r}' w_{in_{1}}^{\sigma_{1}*}(\boldsymbol{r}) w_{jn_{3}}^{\sigma_{2}*}(\boldsymbol{r}') W(\boldsymbol{r},\boldsymbol{r}',\omega) w_{jn_{4}}^{\sigma_{2}}(\boldsymbol{r}') w_{in_{2}}^{\sigma_{1}}(\boldsymbol{r}).
\end{gathered}
\label{hubudef31}
\end{equation}
\newline
Here $w_{in}^{\sigma}(\boldsymbol{r})$ is a maximally localized Wannier function located at site $i$ and spin $\sigma$. 
In this work we only consider the static limit ($\omega=0$). Furthermore, we use Slater parametrization,
\begin{equation}
\begin{gathered}
U_{i}=\frac{1}{(2l+1)^{2}}\sum_{m,m'} U_{im,im',im,im'}^{\sigma_{1},\sigma_{2}}(\omega=0) \quad \text{and} \\
V_{ij}=\frac{1}{(2l+1)^{2}}\sum_{m,m'} U_{im,jm',im,jm'}^{\sigma_{1},\sigma_{2}}(\omega=0).
\end{gathered}
\label{fhuburpa}
\end{equation}
\newline
Here $U_{i}$ is the screened (bare) onsite Coulomb interaction at site $i$ and $V_{ij}$ the screened (bare) inter-site Coulomb 
interaction between sites $i$ and $j$. Note that although the matrix elements of the Coulomb potential are formally 
spin-dependent due to the spin dependence of the Wannier functions, we find that this dependence is negligible in 
practice.

\section{Results and Discussion}
\label{section-4}

\begin{figure}[t]
\begin{center}
\includegraphics[scale=0.45]{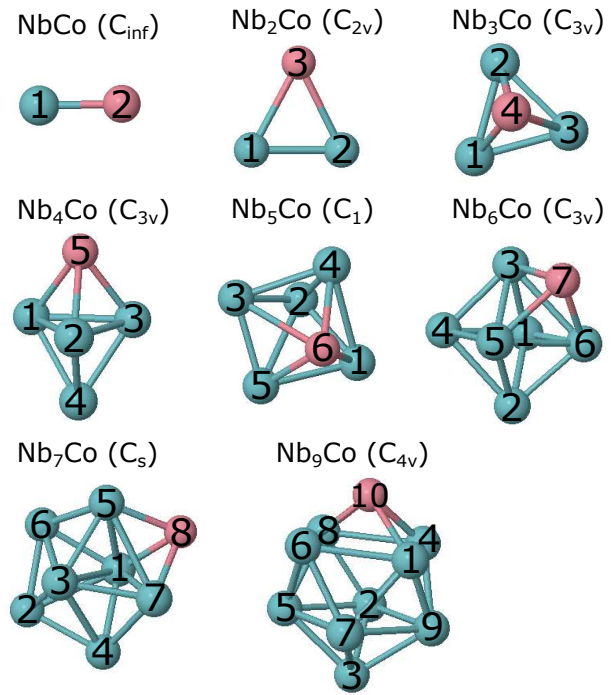}
\end{center}
\vspace{-0.3 cm}
\caption{The geometry of the NbCo to Nb$_{7}$Co and Nb$_{9}$Co clusters. 
Between brackets the point symmetry group of the cluster is indicated. 
The blue spheres correspond to the Nb atoms and the red spheres to the Co atoms. } 
\label{figclus}
\end{figure}

In Fig.\,\ref{figclus} the geometry of the NbCo to Nb$_{7}$Co and Nb$_{9}$Co clusters is depicted. The blue spheres 
correspond to the Nb atoms and the red spheres to the Co atoms. Between brackets the point symmetry group of the 
clusters is indicated. In the following we first address the fully screened (RPA) and partially screened without 
the Co 3\emph{d} $\rightarrow$ 3\emph{d} channel (cRPA) onsite Coulomb interaction matrix elements of the Co 
3\emph{d} electrons for the Nb$_{x}$Co clusters. This provides insight 
into the contribution of the Co 3\emph{d} $\rightarrow$ 3\emph{d} channel to the total screening process. Second, the 
fully screened onsite and inter-site Coulomb interaction matrix elements of the Nb 4\emph{d} 
and Co 3\emph{d} orbitals are investigated. Finally, we make a comparison with pure bcc Nb bulk and investigate the influence of the Nb 4\emph{d} $\rightarrow$ 4\emph{d} channel on the screening of the onsite and inter-site Coulomb interaction of the Nb 4\emph{d} electrons. Note that the partially screened onsite Coulomb interaction is usually referred to as Hubbard $U$ and is what enters effective models, e.g. the Hubbard model.

In Table~\ref{tabuco}, the bare, partially screened without the Co 3\emph{d} $\rightarrow$ 3\emph{d} channel (cRPA) and fully screened (RPA) average onsite Coulomb interaction 
matrix elements of the Co 3\emph{d} electrons are presented. As it is seen the bare interaction is constant as 
function of cluster size, while the partially and fully screened interactions decrease with size. This can be attributed
to the increase of screening channels with increasing cluster size rather than the delocalization of the Wannier functions. 
Note that very similar matrix elements for the onsite bare Coulomb interaction for all clusters reflect the fact that 
the localization of the Wannier functions does not change with increasing the cluster size. Furthermore, the obtained cRPA
and RPA Coulomb matrix elements for the Co 3\textit{d} orbitals are very close to each other, which means that the 
contribution of the Co 3\emph{d} $\rightarrow$ 3\emph{d} channel to the total screening is very small compared to the 
other screening channels (see Table~\ref{tabuco}).

\begin{table}[t]
\begin{center}
\caption{The bare, partially screened without the Co 3\emph{d} $\rightarrow$ 3\emph{d} channel (cRPA) and fully screened (RPA) average onsite Coulomb interaction 
parameters for the Co 3\emph{d} orbitals of the NbCo-Nb$_{7}$Co and Nb$_{9}$Co clusters obtained from \emph{ab-initio} 
calculations.}
\begin{ruledtabular}
\begin{tabular}{lccc}
Cluster & Bare (eV) & cRPA (eV) & RPA (eV)   \\ \hline
NbCo            &  22.2 & 7.9 & 7.7  \\ 
Nb$_{2}$Co & 22.2 & 5.9 & 5.8  \\
Nb$_{3}$Co & 22.3 & 5.6 & 5.5  \\
Nb$_{4}$Co & 22.6 & 5.2 & 5.0  \\
Nb$_{5}$Co & 22.7 & 4.7 & 4.6  \\
Nb$_{6}$Co & 22.7 & 4.4 & 4.3  \\
Nb$_{7}$Co & 22.7 & 4.1 & 4.1  \\
Nb$_{9}$Co & 22.9 & 3.9 & 3.8  \\
\end{tabular}
\end{ruledtabular}
\label{tabuco}
\end{center}
\end{table}

In Table~\ref{tabx} and~\ref{tabxx}, the bare (fourth column) and fully screened (fifth column) 
onsite and inter-site average Coulomb interaction parameters for Nb 4\emph{d} and Co 
3\emph{d} orbitals are presented for the Nb$_{x}$Co clusters. Due to the symmetry of 
some clusters (see Fig.~\ref{figclus}), some Nb atoms are equivalent. In Fig.~\ref{figclus} 
for Nb$_{2}$Co atoms 1 and 2 are equivalent, for Nb$_{3}$Co 1, 2 and 3 are equivalent, 
for Nb$_{4}$Co 1, 2 and 3 are equivalent, for Nb$_{5}$Co there are no equivalent atoms, 
for Nb$_{6}$Co 3, 5 and 6, and 1, 2 and 4 are equivalent, for Nb$_{7}$Co 5 and 7, and 4 
and 6 are equivalent and for Nb$_{9}$Co 1, 4, 6 and 8, and 2, 5, 7 and 9 are equivalent. 
In Table~\ref{tabx} and~\ref{tabxx} only symmetry unequivalent interactions are shown. Further, in 
the first column $U_{1}$ corresponds to the onsite Coulomb interaction of atom 1 and $V_{1,2}$ 
to the inter-site Coulomb interaction between atoms 1 and 2 (see Fig.~\ref{figclus}). The 
second column indicates between what type of atoms this refers and the third column correponds 
to the distance in \AA ~between them. From this table it can be seen that besides for 
the Co 3\emph{d} electrons also for the Nb 4\emph{d} electrons the onsite Coulomb interaction 
is well screened and decreases with cluster size. On the other hand, the inter-site Coulomb 
interaction is much less screened and is more or less constant as a function of interatomic 
distance. This appears to be due to a decrease in the screening as a function of increasing 
interatomic distance. Interestingly for NbCo the inter-site interaction is unscreened, while 
for Nb$_{4}$Co there is even anti-screening present between Nb and Co at an interatomic distance 
of 3.89~\AA. Anti-screening means that the the screened interaction is larger than the bare 
interaction. By considering the effective interaction between two point charges in a medium, 
screening is understood to be due to the response (polarization) of the medium to these charges. 
Similarly anti-screening occurs when the medium is polarized in such a way to increase the bare 
interaction between the two point charges. This situtation is known to occur only for low dimensional 
systems such as carbon nanotubes, nanoribbons, wires, molecules and clusters.\cite{mot1,mot2} 
From a simplistic point of view the two induced point charges can be considered as giving rise 
to point-dipoles at the positions of the polarizable atoms that constitute the medium. Each point-dipole 
produces an electric field and depending on its orientation it either increases or decreases 
the bare electric field coming from the two point charges. Roughly the point-dipoles in between 
the two point charges are oriented to increase, anti-screen, the bare interaction, whereas the 
other surrounding point-dipoles lead to a reduction, screening, of the bare interaction.\cite{mot2} 
Therefore, the occurence of anti-screening crucially depends on the dimensionality of the system and 
distance between the induced point charges. More precisely, for low dimensional systems the ratio 
of the region between the point charges and the rest of the medium is larger.

Anti-screening was also recently found in Fe$_{x}$O$_{y}$ clusters by means of ab-initio 
calculations.\cite{lars3} However, the anti-screening appears to be more pronounced in Fe$_{x}$O$_{y}$ 
clusters than in Nb$_{x}$Co clusters. In order to qualitatively understand this, the microscopic 
point-dipole model can be used. Within this model the atoms of the system are considered as 
classical polarizable point-dipoles. These point-dipoles are then allowed to respond to an 
external perturbation, e.g. induced point charges. From investigations on low-dimensional 
systems by means of this microscopic point-dipole model, it is well established that anti-screening 
delicately depends on the geometry and polarizability of the atoms constituting the system.\cite{mot1,mot2} 
However, in general it is demonstrated that the inter-atomic distance at which anti-screening 
occurs increases with increasing polarizability (see for example Fig.~1.10 of Ref.~\onlinecite{mot2}). 
Further, from for example ab-initio calculations on isolated atoms it is known that the polarizability 
of Nb is larger than that of Fe, Co and O.\cite{atompol} Fe and Co have a similar polarizability, 
which is again larger than that of O. Based on these observations anti-screening in 
Nb$_{x}$Co clusters is expected to occur at larger inter-atomic distances compared to Fe$_{x}$O$_{y}$ 
clusters, which explains why anti-screening is more pronounced in the latter.

\begin{figure}[!t]
\begin{center}
\includegraphics[scale=0.67]{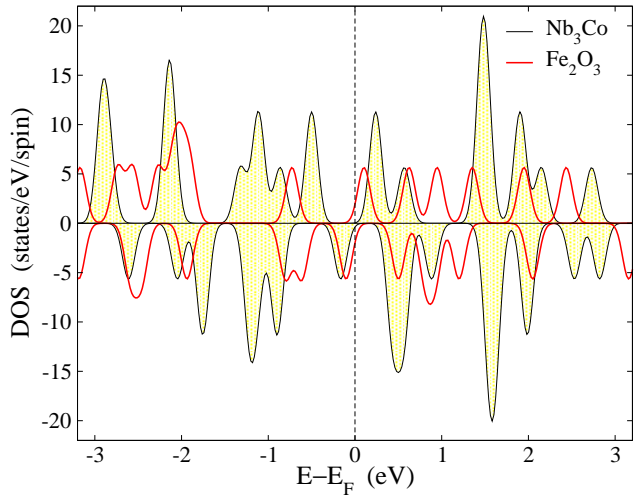}
\end{center}
\vspace{-2.4 cm}
\caption{Calculated spin-resolved total density of states for Fe$_{2}$O$_{3}$ and Nb$_{3}$Co clusters. 
The Fermi energy (chemical potential) is set to zero.}
\label{fig2}
\end{figure}

The discussion above on the difference in anti-screening between Nb$_{x}$Co and  Fe$_{x}$O$_{y}$ clusters 
is based on the microscopic point-dipole model. It is however not clear if these clusters can be modeled 
by a collection of point-dipoles. Therefore, it is instructive to also discuss anti-screening differences 
based on Eq.~(\ref{rpapol1}). It is known that anti-screening only occurs in low-dimensional semiconductors 
and insulators.\cite{mot1,mot2,Louie,Nomura}  As mentioned above, the critical distance for the appearance of 
anti-screening increases with increasing polarizability, which can be traced back to the distribution of the 
occupied and unoccupied electronic states around the Fermi energy (strictly speaking chemical potential 
for the clusters). In Fig.\,\ref{fig2} we present the density of states for Fe$_2$O$_3$ and Nb$_3$Co 
clusters, which is calculated using the Gaussian method  with a broadening parameter of 0.1 eV. 
The polarizability (see Eq.~(\ref{rpapol1})) is inversely proportional to the energy difference between 
occupied and unoccupied states, i.e., the smaller the energy difference the larger the polarizability.  Indeed, as 
seen in Fig.\,\ref{fig2} the Nb$_3$Co cluster has more states around the chemical potential with respect to 
the Fe$_2$O$_3$ cluster despite similar HOMO-LUMO energy gaps of both clusters. As a consequence, the contribution 
of the term between square brackets in Eq.~(\ref{rpapol1}) is larger for the Nb$_3$Co cluster giving rise to
smaller Coulomb matrix elements and absence of anti-screening for inter-site Coulomb interactions. Thus, similar as 
for the microscopic point-dipole model, a small polarization or equivalently polarizability of the system 
is required to observe anti-screening at short distances.

\begin{table}[H]
\begin{center}
\caption{The bare and fully screened (RPA) average Coulomb interaction 
parameters for the Nb 4\emph{d} and Co 3\emph{d} orbitals for the NbCo to Nb$_{5}$Co 
obtained from \emph{ab-initio} calculations. Here $U_{1}$ 
corresponds to the onsite Coulomb interaction of atom 1 and $V_{1,2}$ to the inter-site 
Coulomb interaction between atoms 1 and 2 (see Fig.~\ref{figclus}). The second column 
indicates between what type of atoms this refers and the third column correponds to 
the distance in \AA{ }between them. Note that due the symmetry of some clusters, some 
Nb atoms are equivalent.}
\begin{ruledtabular}
\begin{tabular}{lcccc}
&  & NbCo && \\
U/V & Atom        & Distance (\AA) & Bare (eV) & RPA (eV)   \\ \hline
$U_{1}$ & Nb      & 0              &  11.2 & 7.2    \\ 
$U_{2}$ & Co      & 0              &  22.2 & 7.7  \\
$V_{1,2}$ & Nb-Co & 1.99           & 7.0 & 7.0   \\ \hline
&&&  \\
&  & Nb$_{2}$Co && \\
U/V & Atom & Distance (\AA) & Bare (eV) & RPA (eV)   \\ \hline
$U_{1}$ & Nb & 0 & 10.0 & 5.2  \\ 
$U_{3}$ & Co & 0 & 22.2 & 5.8  \\
$V_{1,2}$ & Nb-Nb & 2.16 & 6.0 & 4.9  \\
$V_{1,3}$ & Nb-Co & 2.33 & 6.1 & 5.0  \\ \hline
&&&  \\
	& & Nb$_{3}$Co && \\
U/V & Atom & Distance (\AA) & Bare (eV) & RPA (eV)   \\ \hline
$U_{1}$ & Nb & 0 & 10.7 & 5.0  \\ 
$U_{4}$ & Co & 0 & 22.3 & 5.5  \\
$V_{1,2}$ & Nb-Nb & 2.40 & 5.6 & 4.5  \\
$V_{1,4}$ & Nb-Co & 2.47 & 5.7 & 4.6  \\ \hline
 &&&  \\
&& Nb$_{4}$Co && \\
U/V & Atom & Distance (\AA) & Bare (eV) & RPA (eV)   \\ \hline
$U_{1}$ & Nb & 0 & 11.0 & 4.6  \\ 
$U_{4}$ & Nb & 0 & 10.8 & 4.5  \\
$U_{5}$ & Co & 0 & 22.6 & 5.0  \\
$V_{1,5}$ & Nb-Co & 2.40 & 6.0 & 4.2  \\
$V_{1,4}$ & Nb-Nb & 2.52 & 5.5 & 4.1  \\
$V_{1,2}$ & Nb-Nb & 2.61 & 5.4 & 4.1  \\
$V_{4,5}$ & Nb-Co & 3.89 & 4.0 & \textbf{4.1}  \\ \hline 
&&&  \\
&& Nb$_{5}$Co && \\
U/V & Atom & Distance (\AA) & Bare (eV) & RPA (eV)   \\ \hline
$U_{1}$ & Nb & 0 & 11.2 & 4.3  \\ 
$U_{2}$ & Nb & 0 & 10.8 & 4.3  \\
$U_{3}$ & Nb & 0 & 11.0 & 4.3  \\
$U_{4}$ & Nb & 0 & 11.2 & 4.3  \\
$U_{5}$ & Nb & 0 & 11.2 & 4.3  \\
$U_{6}$ & Co & 0 & 22.7 & 4.6  \\
$V_{1,6}$ & Nb-Co & 2.27 & 6.3 & 3.9  \\
$V_{4,6}$ & Nb-Co & 2.28 & 6.2 & 3.8  \\
$V_{5,6}$ & Nb-Co & 2.32 & 6.1 & 3.8  \\
$V_{3,5}$ & Nb-Nb & 2.40 & 5.7 & 3.8  \\
$V_{1,2}$ & Nb-Nb & 2.44 & 5.6 & 3.8  \\
$V_{2,3}$ & Nb-Nb & 2.49 & 5.5 & 3.8  \\
$V_{2,4}$ & Nb-Nb & 2.50 & 5.5 & 3.8  \\
$V_{3,4}$ & Nb-Nb & 2.63 & 5.3 & 3.7  \\
$V_{1,5}$ & Nb-Nb & 2.69 & 5.3 & 3.7  \\
$V_{3,6}$ & Nb-Co & 2.81 & 5.1 & 3.8  \\
$V_{1,4}$ & Nb-Nb & 2.92 & 4.9 & 3.7  \\
$V_{2,5}$ & Nb-Nb & 2.99 & 4.8 & 3.7  \\
$V_{2,6}$ & Nb-Co & 3.27 & 4.5 & 3.8  \\
$V_{4,5}$ & Nb-Nb & 3.68 & 4.1 & 3.7  \\
$V_{1,3}$ & Nb-Nb & 3.72 & 4.1 & 3.7  \\
\end{tabular}
\end{ruledtabular}
\label{tabx}
\end{center}
\end{table}

\begin{table}[H]
\begin{center}
\caption{The same as Table~\ref{tabx} for Nb$_{6}$Co,  Nb$_{7}$Co and  Nb$_{9}$Co clusters. Here $U_{1}$ 
corresponds to the onsite Coulomb interaction of atom 1 and $V_{1,2}$ to the inter-site 
Coulomb interaction between atoms 1 and 2 (see Fig.~\ref{figclus} for the geometry of the corresponding clusters).}
\begin{ruledtabular}
\begin{tabular}{lcccc}
&& Nb$_{6}$Co && \\
U/V & Atom & Distance (\AA) & Bare (eV) & RPA (eV)   \\ \hline
$U_{1}$ & Nb & 0 & 11.3 & 4.0  \\ 
$U_{3}$ & Nb & 0 & 11.3 & 4.0  \\
$U_{7}$ & Co & 0 & 22.7 & 4.3  \\
$V_{3,7}$ & Nb-Co & 2.33 & 6.2 & 3.5  \\
$V_{3,4}$ & Nb-Nb & 2.53 & 5.5 & 3.5  \\
$V_{1,2}$ & Nb-Nb & 2.73 & 5.2 & 3.4  \\
$V_{3,5}$ & Nb-Nb & 2.88 & 5.0 & 3.4  \\
$V_{2,3}$ & Nb-Nb & 3.78 & 4.0 & 3.3  \\
$V_{2,7}$ & Nb-Co & 3.91 & 3.9 & 3.4  \\ \hline
&&&  \\
&& Nb$_{7}$Co && \\
U/V & Atom & Distance (\AA) & Bare (eV) & RPA (eV)   \\ \hline
$U_{1}$ & Nb & 0 & 11.3 & 3.9  \\ 
$U_{2}$ & Nb & 0 & 11.2 & 3.9  \\
$U_{3}$ & Nb & 0 & 11.2 & 3.9  \\
$U_{4}$ & Nb & 0 & 11.3 & 3.9  \\
$U_{5}$ & Nb & 0 & 11.4 & 3.9  \\
$U_{8}$ & Co & 0 & 22.7 & 4.1  \\
$V_{5,8}$ & Nb-Co & 2.30 & 6.2 & 3.4  \\
$V_{1,8}$ & Nb-Co & 2.43 & 5.9 & 3.3  \\
$V_{2,4}$ & Nb-Nb & 2.46 & 5.6 & 3.4  \\
$V_{1,4}$ & Nb-Nb & 2.53 & 5.5 & 3.4  \\
$V_{3,5}$ & Nb-Nb & 2.54 & 5.5 & 3.3  \\
$V_{2,3}$ & Nb-Nb & 2.56 & 5.4 & 3.3  \\
$V_{3,4}$ & Nb-Nb & 2.83 & 5.0 & 3.3  \\
$V_{1,5}$ & Nb-Nb & 2.85 & 5.0 & 3.3  \\
$V_{1,2}$ & Nb-Nb & 2.86 & 5.0 & 3.3  \\
$V_{5,7}$ & Nb-Nb & 2.90 & 5.0 & 3.3  \\
$V_{1,3}$ & Nb-Nb & 3.15 & 4.6 & 3.3  \\
$V_{3,8}$ & Nb-Co & 3.78 & 4.1 & 3.3  \\
$V_{4,8}$ & Nb-Co & 3.97 & 3.9 & 3.3  \\
$V_{4,6}$ & Nb-Nb & 3.98 & 3.9 & 3.2  \\
$V_{2,5}$ & Nb-Nb & 4.08 & 3.8 & 3.2  \\
$V_{4,5}$ & Nb-Nb & 4.26 & 3.7 & 3.2  \\
$V_{2,8}$ & Nb-Co & 4.88 & 3.4 & 3.3  \\ \hline
&&&  \\
&& Nb$_{9}$Co && \\
U/V & Atom & Distance (\AA) & Bare (eV) & RPA (eV)   \\ \hline
$U_{1}$ & Nb & 0 & 11.5 & 3.4  \\ 
$U_{2}$ & Nb & 0 & 11.4 & 3.5  \\
$U_{3}$ & Nb & 0 & 11.4 & 3.4  \\
$U_{10}$ & Co & 0 & 22.9 & 3.8  \\
$V_{1,10}$ & Nb-Co & 2.42 & 6.0 & 3.0  \\
$V_{1,7}$ & Nb-Nb & 2.53 & 5.5 & 2.9  \\
$V_{2,3}$ & Nb-Nb & 2.57 & 5.5 & 2.9  \\
$V_{1,4}$ & Nb-Nb & 2.82 & 5.1 & 2.9  \\
$V_{2,5}$ & Nb-Nb & 2.87 & 5.0 & 2.9  \\
$V_{2,10}$ & Nb-Co & 3.94 & 3.9 & 2.9  \\
$V_{1,8}$ & Nb-Nb & 3.99 & 3.9 & 2.9  \\
$V_{2,7}$ & Nb-Nb & 4.05 & 3.8 & 2.8  \\
$V_{1,3}$ & Nb-Nb & 4.11 & 3.8 & 2.8  \\
$V_{1,2}$ & Nb-Nb & 4.22 & 3.7 & 2.8  \\
$V_{3,10}$ & Nb-Co & 4.96 & 3.3 & 2.9  \\ 
\end{tabular}
\end{ruledtabular}
\label{tabxx}
\end{center}
\end{table}

\noindent A similar discussion holds for all other clusters, for 
instance the NbCo cluster has a similar molecular energy level distribution around the chemical potential as Fe$_{3}$O$_{4}$ 
(not shown). Then, anti-screening is expected to occur at similar inter-site distances in these clusters. For 
Fe$_{3}$O$_{4}$ this is expected to occur a bit below $3.4$~\AA~(see Table~I of Ref.~\onlinecite{lars3}), while 
for NbCo indeed just above $3.0$~\AA. Further, although Nb$_{2}$Co and Nb$_{3}$Co show a similar molecular energy 
spectrum around the chemical potential as the Fe$_{4}$O$_{6}$ cluster, anti-screening is not observed, because 
the inter-site distances are too small compared to Fe$_{4}$O$_{6}$. For Nb$_{4}$Co and larger clusters the density of
molecular energy levels around the chemical potential increases and is quite a bit denser than for the Fe$_{x}$O$_{y}$ 
clusters. Therefore, anti-screening in these clusters is only expected for large inter-site distances. For example, 
in Nb$_{4}$Co it occurs at $3.89$~\AA, while for Nb$_{7}$Co at an inter-site Nb-Co distance of $4.88$~\AA~the situation 
is very close to anti-screening.

It is instructive to compare the Nb$_{x}$Co results with pure Nb bulk. In Table~\ref{tabbulk} the bare, 
partially screened (without the Nb 4\emph{d} $\rightarrow$ 4\emph{d} channel), and fully screened onsite 
and inter-site Coulomb interaction matrix elements of the Nb 4\emph{d} electrons are presented. From this 
table it is clear that the inter-site Coulomb interaction in RPA is almost completely screened. This is in 
contrast with what is observed for the clusters (Table~\ref{tabx} and~\ref{tabxx}). Further, the fully 
screened onsite Coulomb interaction is more screened than for the clusters. The important observation 
for pure Nb bulk is that the effective Coulomb interaction is not constant throughout the system. Instead, 
it is localized, i.e. short ranged with a large gradient. Further, the contribution of the Nb 4\emph{d} 
$\rightarrow$ 4\emph{d} channel to the screening can be investigated from Table~\ref{tabbulk}. Both for 
the onsite and inter-site effective interaction this contribution is small, about 1.8~eV and 0.07~eV 
(for the nearest-neighbor interaction), compared to the contribution of about 11~eV and 4.9~eV of 
the other channels. The main screening contribution comes from the 5\emph{s} states, which are present around the Fermi level.

\begin{table}[!b]
\begin{center}
\caption{The bare, partially screened without the Nb 4\emph{d} $\rightarrow$ 4\emph{d} channel (cRPA), and fully screened (RPA) average onsite and inter-site Coulomb interaction parameters for the Nb 4\emph{d} orbitals of pure Nb bulk. Here the 
first column refers to the distance in~\AA{ } between two Nb atoms, i.e. zero corresponds to the onsite 
interaction.} 
\begin{ruledtabular}
\begin{tabular}{lccc}
Distance (\AA)  & Bare (eV) & cRPA (eV) & RPA (eV)   \\ \hline
0     &  13.81 & 2.62   &  0.83 \\ 
2.86  &  5.01  & 0.08   &  0.01 \\
3.30  &  4.35  & 0.04   &  0.00 \\
4.67  &  3.11  & 0.01   &  0.00 \\
5.72  &  2.57  & 0.00   &  0.00 \\
\end{tabular}
\end{ruledtabular}
\label{tabbulk}
\end{center}
\end{table}

\begin{figure}[!t]
\begin{center}
\includegraphics[scale=0.7]{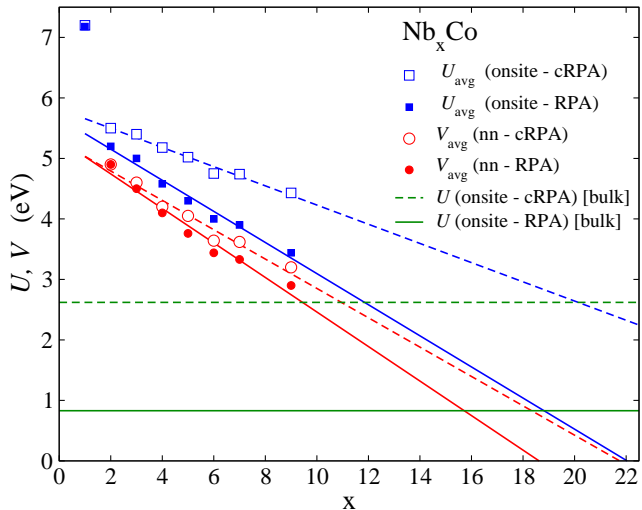}
\end{center}
\vspace{-0.5 cm}
\caption{The averaged partially screened without the Nb 4\emph{d} $\rightarrow$ 4\emph{d} channel and fully 
screened onsite $U_{avg}$ (onsite) and nearest-neighbour inter-site $V_{avg}$ (nn) matrix elements 
between Nb 4\emph{d} electrons as function of cluster size ($x$) for the Nb$_{x}$Co clusters. Here $x$ 
corresponds to the number of Nb atoms in the Nb$_{x}$Co cluster. Ignoring the smallest 
cluster ($x=1$), the blue and red solid and dashed lines represent an extrapolation of the data points. 
The green solid and dashed lines represent respectively the onsite fully screened and partially screened 
Coulomb interaction for pure Nb bulk.} 
\label{fig3}
\end{figure}

For the Nb$_{x}$Co clusters the influence of the Nb 4\emph{d} $\rightarrow$ 4\emph{d} channel 
can be obtained from an inspection of Fig.~\ref{fig3}. Here an average of the partially screened 
(without the Nb 4\emph{d} $\rightarrow$ 4\emph{d} channel) and fully screened onsite and 
nearest-neighbor inter-site Coulomb interaction parameters for the Nb 4\emph{d} orbitals are 
presented as function of cluster size. The cluster size is indicated by $x$, which represents 
the number of Nb atoms in the Nb$_{x}$Co clusters. It appears that the contribution of the Nb 
4\emph{d} $\rightarrow$ 4\emph{d} channel to the screening of the onsite and inter-site effective 
interaction increases with cluster size. For instance for the onsite interaction this contribution 
is about 0.3~eV for Nb$_{2}$Co and becomes about 0.9~eV for the Nb$_{9}$Co cluster. In case of 
the nearest-neighbour inter-site interaction the contribution for Nb$_{2}$Co is almost 0~eV and 
becomes about 0.3~eV for Nb$_{9}$Co. Compared to the contributions of the other channels (see 
Table~\ref{tabx} and~\ref{tabxx} for the unscreened bare values), it can be concluded that the 
contribution of the Nb 4\emph{d} $\rightarrow$ 4\emph{d} channel to the screening is small. 
Namely for the onsite interaction the contribution of the other channels is about 5~eV for 
Nb$_{2}$Co and becomes about 8~eV for Nb$_{9}$Co. In case of the nearest-neighbor inter-site 
interaction this is about 1~eV for Nb$_{2}$Co and 2.5~eV for Nb$_{9}$Co.

It is interesting to obtain insight at what cluster size the behavior of the screened Coulomb interaction 
becomes bulk like. For this purpose Fig.~\ref{fig3} is used again, where the averaged onsite and nearest-neighbor 
inter-site screened and partially screened Coulomb interaction parameters between Nb 4\emph{d} electrons 
are presented as function of cluster size and compared with the pure Nb bulk values of Table~\ref{tabbulk} 
(green dashed and solid lines for the onsite cRPA and RPA interactions, respectively). The blue and red solid 
(dashed) lines correspond to a linear extrapolation of the Nb$_{x}$Co cluster data points for which the smallest 
cluster ($x=1$) is ignored. From these extrapolations it appears that both the averaged onsite and inter-site 
screened and partially screened Coulomb interaction depend in a reasonable approximation linearly on the cluster 
size. At a cluster size of $x=20$ both the onsite and nearest-neighbor inter-site screened (partially screened) 
interaction have reached their corresponding bulk values, i.e. 0.83 eV (2.62 eV) and 0.01~eV (0.08 eV), respectively. 
Therefore, we expect Nb$_{x}$Co clusters with $x$ larger than 20 to have a bulk like behavior.

Finally, we would like to comment on the strength of the electronic correlations in the Nb$_{x}$Co clusters 
and Nb bulk. As shown in Table~\ref{tabx} and~\ref{tabxx} the effective Coulomb interaction is more or less 
constant throughout the clusters. In contrast for Nb bulk it has a strong gradient and is local in nature. 
Although the effective Coulomb interaction in Nb bulk is mainly local, it should not be considered as a 
(strongly) correlated material. For this purpose the band width should also be taken into account. The band 
width is about 7.5~eV~\cite{nb1}, which is much larger than the effective onsite Coulomb interaction of 0.83~eV (see Table~\ref{tabbulk}). Therefore, it should be interpreted as a weakly correlated material and standard DFT is 
expected to provide a good description of the essential physics. This is confirmed by DFT studies on the elastic 
properties, band structure and electron-phonon coupling of Nb bulk, which are in good agreement with experiments.\cite{nb1,nb2,nb3} Due to the almost constant effective Coulomb interaction in the Nb$_{x}$Co clusters, it is also expected that DFT 
should be able to capture the essential physics. This is confirmed by a comparison of the vibrational spectra 
obtained within DFT and experiments.\cite{larsrem} Furthermore, in Ref.~\onlinecite{hong} it is correctly predicted 
within DFT that Nb$_{7}$Co should be non-magnetic. The wrong prediction of Nb$_{5}$Co to be magnetic is probably 
due to the consideration of the wrong geometry (see Ref.~\onlinecite{larsrem}). Besides providing an explanation 
for the success of DFT in these clusters, our results are crucial to select an adequate method for future 
investigations on many-body effects, e.g. quasi-particle life times. For example, intuitively one might expect DFT 
in combination with the dynamical mean-field theory (DMFT)~\cite{dmft} to be suitable for this purpose, because the 
Co atom can be interpreted as an impurity in a Nb$_{x}$ host. Since DMFT only properly treats local correlations, 
while we have demonstrated non-local correlations to be also important, this is not a justified choice. Therefore, 
an extended Hubbard-like model or the consideration of the cluster within multiplet ligand-field theory~\cite{mlft} 
are probably more suitable choices. 

In addition we expect, that due to the almost constant effective interaction in the Nb$_{x}$Co clusters, the observed trends are robust with respect to the choice of the exchange-correlation functional. For example, the local density approximation (LDA)~\cite{lda1} and GGA are expected to perform similar due to the constant interaction, because both methods are derived in the limit of a (nearly) uniform electron gas. As a test we made for all clusters a comparison between the density of states in GGA and LDA. Since they were found to be very similar around the Fermi level, it is indeed expected based on Eq.~\ref{rpapol1} that our results are robust with respect to the choice of the exchange-correlation functional.

\section{Conclusion}
\label{section-5}

We have performed RPA and cRPA calculations to reveal the screening of the Coulomb interaction in Nb$_{x}$Co ($1 \leq x \leq 9$) 
clusters and pure Nb bulk. We have found that in both the clusters and the bulk the onsite Coulomb interaction in RPA
is well screened. On the other hand the inter-site Coulomb interaction is much less screened in the clusters resulting in an almost 
constant interaction throughout the clusters. This is in contrast with pure Nb bulk, where the inter-site Coulomb interaction
in RPA is almost completely screened. Our cRPA calculations have shown that the contribution of the Co 3\emph{d} $\rightarrow$ 
3\emph{d} channel to the total screening process of the onsite Coulomb parameters of the Co 3\emph{d} electrons is negligible. Further, for the clusters investigated the contribution of the Nb 4\emph{d} $\rightarrow$ 4\emph{d} channel to effective onsite 
and inter-site Coulomb parameters of the Nb 4\emph{d} electrons appears to be small compared to that of the total screening
contribution. Based on our findings we expect both for the Nb$_{x}$Co clusters and Nb bulk that correlation effects play a 
minor role and that standard DFT is able to capture the essential physics. For the clusters this is due to the almost constant
effective Coulomb interaction and for the bulk due to the band width being much larger than the essentially local effective 
Coulomb interaction. Finally, it has been found that both the onsite and inter-site Coulomb interaction parameters decrease 
in a reasonable approximation linearly with cluster size and for Nb$_{x}$Co clusters having more than 20 Nb atoms a transition 
from 0D to 3D screening is expected to take place.

\subsection*{Acknowledgements}
The Nederlandse Organisatie voor Wetenschappelijk Onderzoek (NWO) and SURFsara 
are acknowledged for the usage of the LISA supercomputer and their support. L.P. and
M.I.K. acknowledges a support by European ResearchCouncil (ERC) Grant No. 338957. E. S. and I. M. greatly acknowledge the funding provided by the European Union (EFRE).



\end{document}